\documentclass[aps,pre,twocolumn,amsmath,superscriptaddress]{revtex4-2}
\usepackage{silence}
\usepackage{times}
\usepackage{color}
\usepackage{hyperref}
\usepackage{url}
\usepackage{soul}
\usepackage{graphicx}
\usepackage{amsmath}
\usepackage{subfigure}
\usepackage{xcolor}
\usepackage{bibunits}
\usepackage{amssymb}
\usepackage{latexsym}
\DeclareGraphicsExtensions{.png,.eps}
\usepackage{float}
\usepackage{appendix}
\usepackage{etoolbox}
\usepackage{physics}
\usepackage{tikz}
\usepackage[normalem]{ulem}
\usetikzlibrary{calc,positioning}

\newcommand{\kket}[1]{|#1\rangle\!\rangle}

\hypersetup{
    colorlinks = true,
    citecolor = {blue},
    linkcolor = {purple},
}

\begin{document}

\newcommand{\sutdphys}{Science, Mathematics and Technology Cluster, Singapore
University of Technology and Design, 8 Somapah Road, 487372 Singapore}
\newcommand{\sutdepd}{EPD Pillar, Singapore University of Technology and Design, 8 Somapah Road, 487372 Singapore}
\newcommand{\sutdistd}{ISTD Pillar, Singapore University of Technology and Design, 8 Somapah Road, 487372 Singapore}
\newcommand{\cqt}{Centre for Quantum Technologies, National University of Singapore 117543, Singapore} 
\newcommand{\majulab}{MajuLab, CNRS-UNS-NUS-NTU International Joint Research Unit, UMI 3654, Singapore}

\title{Lindbladian spectral statistics beyond no-jump Hamiltonians: roles of recycling and Liouville-space structure}

\author{Dingzu Wang}
\affiliation{\sutdphys}
\affiliation{\cqt}
\author{Hao Zhu}
\email[Corresponding author: ]{zhuhao6590@buaa.edu.cn}
\affiliation{School of Physics, Beihang University,100191,Beijing, China}
\author{Guo-Feng Zhang}
\email[Corresponding author: ]{gf1978zhang@buaa.edu.cn}
\affiliation{School of Physics, Beihang University,100191,Beijing, China}

\author{Dario Poletti}
\email[Corresponding author: ]{dario\_poletti@sutd.edu.sg}
\affiliation{\sutdphys}\affiliation{\sutdepd}
\affiliation{\cqt}
\affiliation{\majulab}

\begin{abstract}
Spectral statistics probe integrability versus chaos and have recently been extended to Markovian open quantum systems described by Lindbladians, whose quantum-trajectory unraveling decomposes the evolution into no-jump dynamics generated by an effective non-Hermitian Hamiltonian and recycling jumps.
In this work, we perform spectrum-statistics diagnostics for Lindbladians and their effective non-Hermitian Hamiltonians. 
We show that recycling processes, symmetry constraints, and the Liouville-space structure crucially shape the spectral correlations. 
In particular, we identify a family of spectrally separable Lindbladians whose spectra exhibit robust Poisson statistics, despite the effective non-Hermitian Hamiltonian varying from Poisson to strongly correlated complex-spectrum statistics.
Our work establishes a unified spectral-statistics characterization for Lindbladians and their associated effective non-Hermitian Hamiltonians, deepening our understanding of spectral properties in open many-body systems. 
\end{abstract}

\maketitle

\section{Introduction}

While quantum dynamics is often idealized as unitary, non-unitary evolution is ubiquitous because realistic systems inevitably interact with their environment.
To describe such environmental effects, the theory of open quantum systems~\cite{2007Breuer,2000Weiss,2012Rivas} provides a standard framework to account for dissipation and decoherence.
In the Markovian (i.e., memoryless) regime, the evolution of an open quantum system is governed by the Gorini-Kossakowski-Sudarshan-Lindblad master equation~\cite{1976JMPGorini,1976CMPLindblad}.
The generator of this master equation, the Lindbladian, underlies relaxation toward steady states and sets the characteristic dynamical timescales~\cite{2007Breuer,1987LNP286A, LandiSchaller2022}.

In recent years, non-Hermitian systems have attracted much attention, for instance, in the study of Lindbladian dynamics of integrable~\cite{2008NJPh10d3026P,2013PhysRevLett110047201,2016PhysRevLett117137202,2017PhysRevX7041015,2018PhysRevLett120090401,2019PhysRevLett122010401} and chaotic~\cite{2020JPhA53D5303S,PhysRevLett123140403,2019PhysRevLett123234103,2019JPhA52V5302C} open quantum systems, localization~\cite{hamazaki2022lindbladianmanybodylocalization,PhysRevLett123090603, XuPoletti2018, XuPoletti2021, VakulchykDenisov2018}, and thermalization~\cite{PhysRevB100045112,2025PhysRevLett134180405,qy19nc4r}. 
More concretely, from the quantum trajectory perspective~\cite{1992PhysRevLett68580,1992PhysRevA454879,1992PhysRevA464382,1993Molmer93,1998RevModPhys70101,2014AdPhy6377D}, Lindbladian dynamics can be unraveled into stochastic trajectories consisting of random quantum jumps interspersed with no-jump segments generated by an effective non-Hermitian Hamiltonian (NHH).
This conditioned description is widely used in studies of continuously monitored open systems~\cite{2014PhysRevX4041001,2019PhysRevLett123123601,2021PhysRevLett126077201}, where postselection on no-jump trajectories makes the effective NHH directly relevant for the conditional dynamics. 
However, it is not obvious whether this dynamical connection carries over to the spectral correlations 
between the Lindbladian and its associated effective NHH.

The study of spectral statistics is of fundamental importance in theoretical physics, owing to its universality and its role as a robust diagnostic of integrability versus chaos~\cite{2004Mehta,1984PhRvL521B,PhysRevA432046,PhysRevE50888,2010qscbookH}.
In closed (Hermitian) systems, spectral statistics distinguish integrability from chaos, 
with Poisson statistics for integrable models, characterized by uncorrelated levels, 
and Wigner–Dyson statistics for chaotic ones, exhibiting level repulsion as predicted by random-matrix theory (RMT)~\cite{CasatiGuarnieri1980, BohigasSchmit1984, 1998PhR299189G}.
For open quantum systems, various extensions of spectral statistics analysis to the complex spectrum of the Lindbladians have been proposed~\cite{1988PhRvL611899G,2019PhRvL123y4101A,PhysRevLett123140403,2020PhRvX10b1019S,2020PhRvR2b3286H,2023PRXQ4d0312K}.
Recent work has also questioned the extent to which complex-spectrum statistics in dissipative quantum systems provide a universal proxy for classical long-time chaos, emphasizing the distinction between transient and steady-state chaos~\cite{Villasenor2024, Villasenor2025, Mondal2026}.
In this context, complex spectral statistics have become a broadly useful probe of 
spectral features
in open and non-Hermitian many-body systems~\cite{2020PhysRevLett124100604,2021PhysRevLett127170602,2022ScPC526R,PhysRevLett128190402,2022PhysRevX12021040,2023PhysRevLett130010401, 2023PhysRevD107066007,2019PhysRevX9041015,2019PhysRevB99235112,2020PhysRevLett124040401,PhysRevX13031019,PRXQuantum4030328, LeePoletti2022}. 
This raises the question of whether the spectral statistics of the Lindbladian reflect those of the associated effective NHH and, if not, what role is played by the quantum-jump (recycling) term.

In this work, we 
explore this question
by characterizing the spectral statistics of Lindbladians and their no-jump effective non-Hermitian Hamiltonians within a unified set of complex spectrum diagnostics.
We map out the four possible combinations of 
spectral statistics
for the two generators within the diagnostic framework and demonstrate each in paradigmatic dissipative spin chains, highlighting how the recycling contribution and the coherent Hamiltonian structure can reshape the spectral correlations.
We then focus on a class of Lindbladians describing $U(1)$-symmetric systems with local damping that we refer to as \textit{spectrally separable}, i.e., for which the spectrum admits an exact construction from the associated effective NHH.
Within this family, the Lindbladian exhibits robust Poisson statistics in the complex spectrum despite the associated effective NHH varying between strongly correlated and Poisson-like behavior.
In the special case of spatially uniform damping, we uncover a distinct phenomenology in which the level statistics can resemble real-spectrum Poissonian behavior, even though the Lindbladian itself has a genuinely complex spectrum. 

The remainder of this paper is organized as follows.
In Sec.~\ref{sec:SecII}, we introduce the Lindbladian and the associated no-jump effective non-Hermitian Hamiltonian, formulate the ladder representation used for symmetry resolution, and summarize the spectral diagnostics employed throughout.
In Sec.~\ref{sec:SecIII}, we present the main correspondence scenarios by analyzing paradigmatic dissipative spin chains, and demonstrate how correspondence or breakdown can arise under different dissipative channels. 
In Sec.~\ref{sec:SecIV}, we identify a broad class of systems in which the Lindbladian exhibits Poissonian statistics independently of the statistics of the non-Hermitian Hamiltonian. 
Finally, Sec.~\ref{sec:SecV} provides a summary and an outlook.

\section{Spectral analysis scheme}\label{sec:SecII}

In this section, we first introduce the Lindbladian dynamics and its effective non-Hermitian Hamiltonian, and then present the ladder representation and spectral diagnostics used throughout this work.

\subsection{Lindbladian and no-jump effective Hamiltonian}

We consider Markovian open quantum systems whose density matrix evolves according to the Gorini-Kossakowski-Sudarshan-Lindblad master equation~\cite{1976JMPGorini,1976CMPLindblad,2007Breuer,2012Rivas}, i.e., $\dot{\rho}=\mathcal{L}[\rho]$, where the Lindbladian superoperator $\mathcal{L}$ is given by

\begin{align}\label{eq:masterEquation}
    \mathcal{L} [\rho]
    &= 
    - \mathrm{i}[H, \rho] + \sum_{i}  ( L_{i} \rho L^{\dagger}_{i} - \frac{1}{2} \{L_{i}^{\dagger} L_{i}, \rho\} ),
\end{align}
with $H$ the system Hamiltonian, and $ \{L_i\} $ the jump operators modeling the system-environment coupling. We set $\hbar=1$ throughout.

It is often convenient to rewrite the Lindbladian in a form that separates the conditional no-jump evolution from the recycling processes,
\begin{align}\label{eq:Lindblad_Heff}
    \mathcal{L}[\rho]
    =
    - \mathrm{i}\left(
        H_\mathrm{eff} \rho - \rho H_\mathrm{eff}^\dagger
    \right)
    + \sum_i L_i \rho L_i^\dagger,
\end{align}
where the effective non-Hermitian Hamiltonian is defined as
\begin{align}\label{eq:NHH}
    H_\mathrm{eff}
    =
    H - \frac{\mathrm{i}}{2}\sum_i L_i^\dagger L_i.
\end{align}

Within the quantum-trajectory picture, the non-Hermitian evolution generated by $H_{\rm eff}$ describes the conditioned dynamics between quantum jumps, while the recycling term $\sum_i L_i \rho L_i^\dagger$ accounts for stochastic jump events.
Ensemble averaging over such trajectories reproduces the Lindbladian dynamics.

\subsection{Ladder representation and symmetries}\label{sec:SecII.B}

\begin{figure}[t]
    \centering
    \pgfdeclarelayer{bg}
    \pgfsetlayers{bg,main}
    \begin{tikzpicture}[scale=0.9] 

    \def\Na{3} 
    \def\Nt{6} 
    \def\sp{1.2} 

    \foreach \x in {1,...,\Nt} {
        \node[circle,draw,fill=white] (k\x) at (\sp*\x,1) {};
        \node[circle,draw,fill=white] (b\x) at (\sp*\x,0) {};
    }

    \pgfmathtruncatemacro{\Nend}{\Na-1}
    \foreach \x in {1,...,\Nend} {
        \pgfmathtruncatemacro{\xnext}{\x+1}
        \draw[line width=0.8pt] (k\x) -- (k\xnext);
        \draw[line width=0.8pt] (b\x) -- (b\xnext);
    }
    \pgfmathtruncatemacro{\Nstart}{\Na+1}
    \pgfmathtruncatemacro{\Nend}{\Nt-1}
    \foreach \x in {\Nstart,...,\Nend} {
        \pgfmathtruncatemacro{\xnext}{\x+1}
        \draw[line width=0.8pt] (k\x) -- (k\xnext);
        \draw[line width=0.8pt] (b\x) -- (b\xnext);
    }

    \def\midsp{0.3}
    \draw[line width=0.8pt] (k\Na) -- ({\Na*\sp+\midsp}, 1);
    \draw[line width=0.8pt] (b\Na) -- ({\Na*\sp+\midsp}, 0);
    \pgfmathtruncatemacro{\xnext}{\Na+1}
    \draw[line width=0.8pt] ({\xnext*\sp-\midsp}, 1) -- (k\xnext);
    \draw[line width=0.8pt] ({\xnext*\sp-\midsp}, 0) -- (b\xnext);
    \node at ({(\Na+0.5)*\sp}, 1) {$\cdots$};
    \node at ({(\Na+0.5)*\sp}, 0) {$\cdots$};

    \foreach \x in {1,...,\Nt} {
        \draw[red, line width=0.8pt] (k\x) -- (b\x);
    }



    \begin{pgfonlayer}{bg}
    \fill[pink!20, rounded corners=6pt]
    (-1,0.7) rectangle ({\Nt*\sp+0.4}, 1.6);
    \fill[blue!9, rounded corners=6pt]
    (-1,-0.64) rectangle ({\Nt*\sp+0.4}, 0.3);
    \end{pgfonlayer}

    \foreach \x in {1,...,\Na} {
        \node[above=4pt] at (k\x) {$\x$};
    }
    \foreach \x in {1,...,\Na} {
        \node[below=4pt] at (b\x) {$\x'$};
    }
    \pgfmathtruncatemacro{\Nstart}{\Na+1}
    \pgfmathtruncatemacro{\Nlastm}{\Nt-1}
    \foreach \x in {\Nstart,...,\Nlastm} {
        \pgfmathtruncatemacro{\i}{\Nt-\x}
        \node[above=4pt] at (k\x) {$L-\i$};
        \node[below=4pt] at (b\x) {$(L-\i)'$};
    }
    \node[above=4pt] at (k\Nt) {$L$};
    \node[below=4pt] at (b\Nt) {$L'$};

    \node[anchor=east] at (0.9,1.4) {ket leg:};
    \node[anchor=east] at (0.9,-0.4) {bra leg:};

    \node[anchor=east] at (0.9,0.9) {$-\,\mathrm{i}\,H_{\rm eff}\otimes\mathbb{I}$};
    \node[anchor=east] at (0.9,0.1) {$+\,\mathrm{i}\,\mathbb{I}\otimes H_{\rm eff}^{*}$};

    \node[red] at (\Nt*\sp+0.8,0.7) {recycling};
    \node[red] at (\Nt*\sp+0.8,0.3) {$L_i \otimes L_{i}^{*}$};

    \end{tikzpicture}

    \caption{Depiction of the ladder representation of the Lindbladian $\mathbb{L}$. The upper (lower) leg corresponds to the ket (bra) Hilbert space. The no-jump part acts independently on the two legs as $-\mathrm i\,H_{\rm eff}\otimes\mathbb I$ and $+\mathrm i\,\mathbb I\otimes H_{\rm eff}^*$, while the recycling term generates inter-leg couplings on each rung, $L_i\otimes L_i^*$. Here $i$ labels the site on the ket leg corresponding to site $i'$ on the bra leg. 
    }
    \label{fig:ladder}
\end{figure}
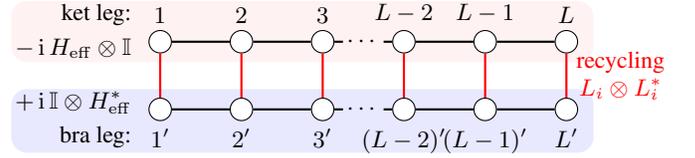

To extract spectral signatures associated with integrable or chaotic behavior, it is essential to resolve all relevant symmetries.
For Lindbladians, this task is more subtle than for isolated Hamiltonians, 
since the generator acts in a doubled Hilbert space and symmetries acting on ket and bra degrees of freedom can be intertwined.
Hence, we interpret the Lindbladian superoperator as a non-Hermitian operator that acts on a \textit{ladder} structure, 
which provides a physically transparent representation of the doubled Hilbert space,
and allows symmetry analysis
to be carried out in close analogy with ordinary Hamiltonian systems 
(see Fig.~\ref{fig:ladder}).
The ladder terminology is used because the models studied below are one-dimensional; in higher dimensions the same vectorization gives two layers of the original lattice.
In our analysis, this representation organizes the ket/bra charge sectors, the leg-exchange operation, and the recycling terms as inter-leg couplings.

This ladder formulation corresponds to working in Liouville space, where the density matrix is vectorized as
\(
\kket{\rho} = \sum_{m,n} \rho_{mn} \kket{m,n},
\)
and the master equation takes the form
\(
(d/dt) \kket{\rho} = \mathbb{L}\kket{\rho}
\) (see, e.g., \cite{LandiSchaller2022}).
In this formulation, the Lindbladian $\mathbb{L}$ acts as a linear operator on Liouville space, i.e., a doubled Hilbert space spanned by the ket and bra degrees of freedom, and takes the explicit form
\begin{equation}
\mathbb{L}
=
-\,\mathrm{i}\bigl(H_{\rm eff}\otimes\mathbb{I}-\mathbb{I}\otimes H_{\rm eff}^*\bigr)
+\sum_i L_i\otimes L_i^* .
\end{equation}
In this vectorized representation, the ladder structure is naturally revealed, since the no-jump evolution generated by $H_{\rm eff}$ acts independently on the two legs, while the recycling term $L_i \rho L_i^\dagger$ gives rise to inter-leg couplings.
In the following, we discuss how $U(1)$ and spatial symmetries are manifest in Liouville space, and how they can be exploited for symmetry considerations. 

When the Lindbladian satisfies $[\mathbb L,\mathcal M_p]=0$, with
$\mathcal M_p = \sum_{i=1}^{L}\sigma_i^z + \sum_{i=1'}^{L'}\sigma_i^z$, 
the generator $\mathbb L$
admits a global $U(1)$ symmetry in Liouville space.
The sum runs over $2L$ sites in the ladder representation, and $\mathcal{M}_p$ is given by the sum of the magnetizations associated with each leg. 
In addition to this total magnetization symmetry, the ladder representation also clearly manifests a second $U(1)$ symmetry associated with the magnetization difference between the two legs. 
For certain Lindbladians, this quantity is conserved, generated by
\(
    \mathcal M_d
    =
    \sum_{i=1}^{L}\sigma_i^z
    -
    \sum_{i=1'}^{L'}\sigma_i^z,
\)
such that $[\mathbb L,\mathcal M_d]=0$.
The generators $\mathcal M_p$ and $\mathcal M_d$ 
provide a natural
organization of conserved charges in Liouville space
and connect to the notions of strong and weak symmetries in open quantum systems~\cite{2012NJPh14g3007B}.
When both symmetries are present, 
the Lindbladian
can be block-diagonalized into
$(m_p,m_d)$
sectors. 
The allowed values are constrained by
$m_p=m_{\rm ket}+m_{\rm bra}$
and
$m_d=m_{\rm ket}-m_{\rm bra}$,
where $m_{\rm ket}$ and $m_{\rm bra}$ denote the magnetizations on the ket and bra legs, respectively.
As a consequence, 
$m_p$ and $m_d$ 
must be either both even or both odd.

We now turn to spatial symmetries of the Lindbladian.
For systems with spatially uniform Hamiltonians and homogeneous dissipation, the ladder representation makes the spatial reflection symmetry $P_x$ along the chain direction manifest in Liouville space, allowing it to be resolved in the same way as an $x$-direction spatial inversion in lattice systems.
In this form, $P_x$ can also be combined naturally with the $U(1)$ symmetries discussed above.

By contrast, reflection along the $y$-direction of the ladder, denoted as $P_y$, does not generically commute with the Lindbladian.
Nevertheless, it plays a distinguished role in the ladder representation,  as it allows one to choose a particularly convenient basis in the Liouville space.
Specifically, one can work in a basis formed by states that are symmetric and antisymmetric under the ladder reflection $P_y$,
\begin{equation}
    \begin{aligned}
        \kket{\phi^{\rm sym}_{mn}}
        &=
        \frac{1}{\sqrt{2}}
        \bigl(\kket{m,n} + P_y\kket{m,n}\bigr), \\
        \kket{\phi^{\rm asym}_{mn}}
        &=
        \frac{\mathrm i}{\sqrt{2}}
        \bigl(\kket{m,n} - P_y\kket{m,n}\bigr),
    \end{aligned}
\end{equation}
for $m\neq n$, where $\kket{m,n}$ denotes the Liouville-space basis introduced above through vectorization, and $P_y \kket{m,n} = \kket{n,m}$ acts by exchanging the two legs of the ladder.
For $m=n$, $\kket{\phi^{\rm sym}_{mm}}$ reduces to $\kket{m,m}$.
In this $P_y$-(anti)symmetric basis, the Liouville-space vector $\kket{\rho}$ is parametrized by purely real coefficients, corresponding to the real and imaginary parts of the density matrix elements, as a direct consequence of its Hermiticity.
Since the Lindblad master equation preserves Hermiticity, this real structure is invariant under time evolution, implying that the Lindbladian is represented by a purely real matrix in this basis.

An equivalent way to express this structure is in terms of an antiunitary symmetry of the Lindbladian.
The antiunitary operator $\mathcal T$ is defined as $\mathcal T \kket{A} = \kket{A^\dagger}$, which in the ladder representation corresponds to $\mathcal T=P_yK$, where $K$ denotes complex conjugation.
Hermiticity preservation implies
\(
[e^{t\mathcal L}(\rho)]^\dagger = e^{t\mathcal L}(\rho^\dagger)
\) for any density operator $\rho$ \footnote{Here, by $e^{t\mathcal L}(\rho)$ we mean the evolution of the density matrix $\rho$ for a time $t$ using Eq.~(\ref{eq:masterEquation})}.
In vectorized form, this implies
\(
\mathcal T e^{t\mathbb L} \kket{\rho} = \kket{[e^{t\mathcal L}(\rho)]^\dagger} = \kket{e^{t\mathcal L}(\rho^\dagger)} = e^{t\mathbb L}\mathcal T\kket{\rho} 
\)
for all $\kket{\rho}$, and hence
\begin{equation}
    \mathcal T e^{t\mathbb L}\mathcal T^{-1} = e^{t\mathbb L},
    \qquad
    \mathcal T \mathbb L \mathcal T^{-1} = \mathbb L .
\end{equation}
Resolving this time-reversal symmetry~\cite{PRXQuantum4030328} in the $P_y$-(anti)symmetric basis, 
leads directly to a purely real matrix representation of the Lindbladian, improving numerical efficiency.

\subsection{Spectral diagnostics}

Spectral properties of
open and non-Hermitian systems 
require diagnostics beyond the standard Hermitian level statistics. 
We therefore use a set of complementary indicators 
that have been developed recently for complex spectra, 
and apply them consistently to both the Lindbladian $\mathbb L$ 
and its effective NHH $H_{\rm eff}$.

As a basic probe of spectral statistics, 
we use the nearest-neighbor level-spacing distribution 
of the complex eigenvalues of $\mathbb{L}$ and $ H_{\text{eff}} $. 
The distribution $ P(s) $ of spacings 
$ s_{n} = \min_{n\neq n'} |E_{n}- E_{n'}| $ 
in the complex plane 
can show very different behavior. 
On one hand,
where the eigenvalues are effectively uncorrelated in the complex plane, 
$ P(s) $ is well described by 
the two-dimensional (2D) Poisson distribution~\cite{1988PhRvL611899G,2019PhRvL123y4101A}
\begin{align}
    P_\text{Poi}(s) = \frac{\pi}{2} s e^{ - \pi s^{2} /4}.
\end{align}
On the other hand, 
the spacings are expected to follow 
the Ginibre ensemble of complex non-Hermitian Gaussian random matrices (GinUE), 
with the Ginibre distribution $P_{\text{Gin}}(s)$ 
given by~\cite{1965JMP6440G,1999PhRvL83484M,2010qscbookH,2020PhRvR2b3286H}
\begin{align}\label{eq:GinUE}
    P_{\text{Gin}}(s) 
    = c \lim_{M \to \infty} 
        \left( 
            \prod_{m = 1}^{M - 1} 
                e_{m}(s^{2}) e^{ - s^{2}} 
        \right) 
        \sum_{m = 1}^{M - 1} 
            \frac{2s^{2m + 1}}{m!\; e_{m}(s^{2})}
\end{align}
with $e_{m}(s^{2}) = \sum_{l = 0}^{m} s^{2l}/l!$ and $c \simeq 1.1429$. 

For non-Hermitian matrices, complex conjugation and transposition are distinct symmetry constraints. 
For a generic non-Hermitian matrix $A$, 
ordinary time-reversal symmetry (TRS) 
is a complex-conjugation constraint,
$\mathcal{T}A\mathcal{T}^{-1}=A$,
whereas the transpose-type symmetry
$\mathcal{C}A^\dagger\mathcal{C}^{-1}=A$ is denoted TRS$^\dagger$, 
with $\mathcal{T}$ and $\mathcal{C}$ antiunitary.
Only the latter modifies the generic complex bulk spacing statistics, leading to class AI$^\dagger$ when $\mathcal{C}^2=+1$ and class AII$^\dagger$ when $\mathcal{C}^2=-1$. In the absence of such a transposition constraint the relevant random-matrix reference is class A, represented by GinUE~\cite{2020PhRvR2b3286H,PhysRevE.106.014146,
PhysRevResearch.7.013098}.

In practice, we unfold the complex spectrum by rescaling the nearest-neighbor distances with the local spectral density.
For each eigenvalue $E_n$, we estimate the local density $\overline{\rho}_n=k/(\pi[d_n^{(k)}]^2)$, where $d_n^{(k)}$ is the distance to its $k$th-nearest neighbor, and define the unfolded spacing as $\widetilde{s}_n=s_n\sqrt{\overline{\rho}_n}$, with $k=30$~\cite{PhysRevA.108.043301}.
The spacings entering $P(s)$ are finally normalized to unit mean.
To suppress edge effects, we retain only eigenvalues whose real and imaginary parts lie within the central $20\%$ of their respective spectral ranges~\cite{PhysRevLett123090603}, while evaluating the nearest neighbors and local density from the complete symmetry-resolved spectrum~\cite{PhysRevE.106.014146,PhysRevResearch.7.013098}.

Another useful spectral diagnostic, which does not require spectral unfolding, is the complex spacing ratio (CSR), as introduced in Ref.~\cite{2020PhRvX10b1019S}    
\begin{align}
    z_{n} = \frac{
                E_{n}^{\mathrm{NN}} - E_{n}
            }{
                E_{n}^{\mathrm{NNN}} - E_{n}
            },
\end{align}
where $E_{n}^{\mathrm{NN}}$ and $E_{n}^{\mathrm{NNN}}$ denote, respectively, 
the nearest- and next-nearest-neighbors of $E_{n}$ 
with respect to the distance in the complex plane. 
By construction, all complex spacing ratios lie inside the unit disk 
and encode both radial and angular information about the complex spectrum. 
For spectra with weak correlations in the complex plane, 
the CSR distribution is approximately uniform over the unit disk. 
By contrast, 
for spectra exhibiting level repulsion,
the CSR distribution shows a reduced probability density at small radii and a nonuniform angular profile.

In addition to complex eigenvalues, 
we also consider the singular-value statistics of $\mathbb L$ and $H_\text{eff}$, 
following Ref.~\cite{2023PRXQ4d0312K}, 
which enables the use of real-spectrum chaos diagnostics familiar from Hermitian systems.
Given a generic non-Hermitian matrix $ A $  (here $A=\mathbb L$ or $H_{\rm eff}$), 
we obtain its singular values 
from the singular-value decomposition $A = U \mathcal{S} V^\dagger$, 
where $\mathcal{S} = \mathrm{diag}(s_1,\ldots,s_N)$ collects the singular values $s_n$. 
Equivalently, the same set $\{s_n\}$ can be obtained 
via the Hermitization method~\cite{1997NuPhB504579F} 
as the positive square roots of the eigenvalues 
of the Hermitian matrices $A^{\dagger}A$ or $AA^{\dagger}$.
In contrast to eigenvalues, 
singular values are always real and nonnegative even for non-Hermitian matrices.
We characterize spectral statistics using the singular-value spacing ratio 
for the ordered set $\{s_n\}$,
\begin{align}
    r_n = 
    \min\!\left( 
        \frac{s_{n+1}-s_n}{s_n - s_{n-1}},\, 
        \frac{s_n - s_{n-1}}{s_{n+1}-s_n} 
    \right),
\end{align}
in analogy with spacing ratios of real eigenvalues 
in Hermitian random-matrix theory~\cite{2007PhRvB75o5111O,2013PhRvL110h4101A}.
For spectra with weak correlations,
the ratios are expected to follow Poisson statistics, 
with an average value $ \langle r \rangle \simeq 0.3863 $.
For spectra with level repulsion, 
the spacing-ratio distribution coincides with 
that of the corresponding Hermitian Gaussian random-matrix ensemble 
in the appropriate symmetry class, 
with the associated class-dependent average value $\langle r\rangle$ 
summarized in Ref.~\cite{2023PRXQ4d0312K}. 
In this work, we analyze the traceless operators 
$\mathbb L - (\mathrm{tr}\mathbb L/\mathrm{tr}\,\mathbb{I})\mathbb{I}$ 
and $H_\mathrm{eff} - (\mathrm{tr}H_\mathrm{eff}/\mathrm{tr}\,\mathbb{I})\mathbb{I}$.

We note that, while singular-value statistics provide convenient Hermitian-like diagnostics for non-Hermitian matrices, recent works have pointed out that SVD-based spectral correlations may disagree with eigenvalue-based diagnostics of complex spectra in certain settings~\cite{2025PhysRevD111L101904,2026arXiv260207349Y}.
Accordingly, throughout this work, we use singular-value statistics as a complementary probe alongside eigenvalue-based diagnostics ($P(s)$ and CSR).

\begin{figure*}[tbp]
    \centering
    \includegraphics[width=0.9\linewidth]{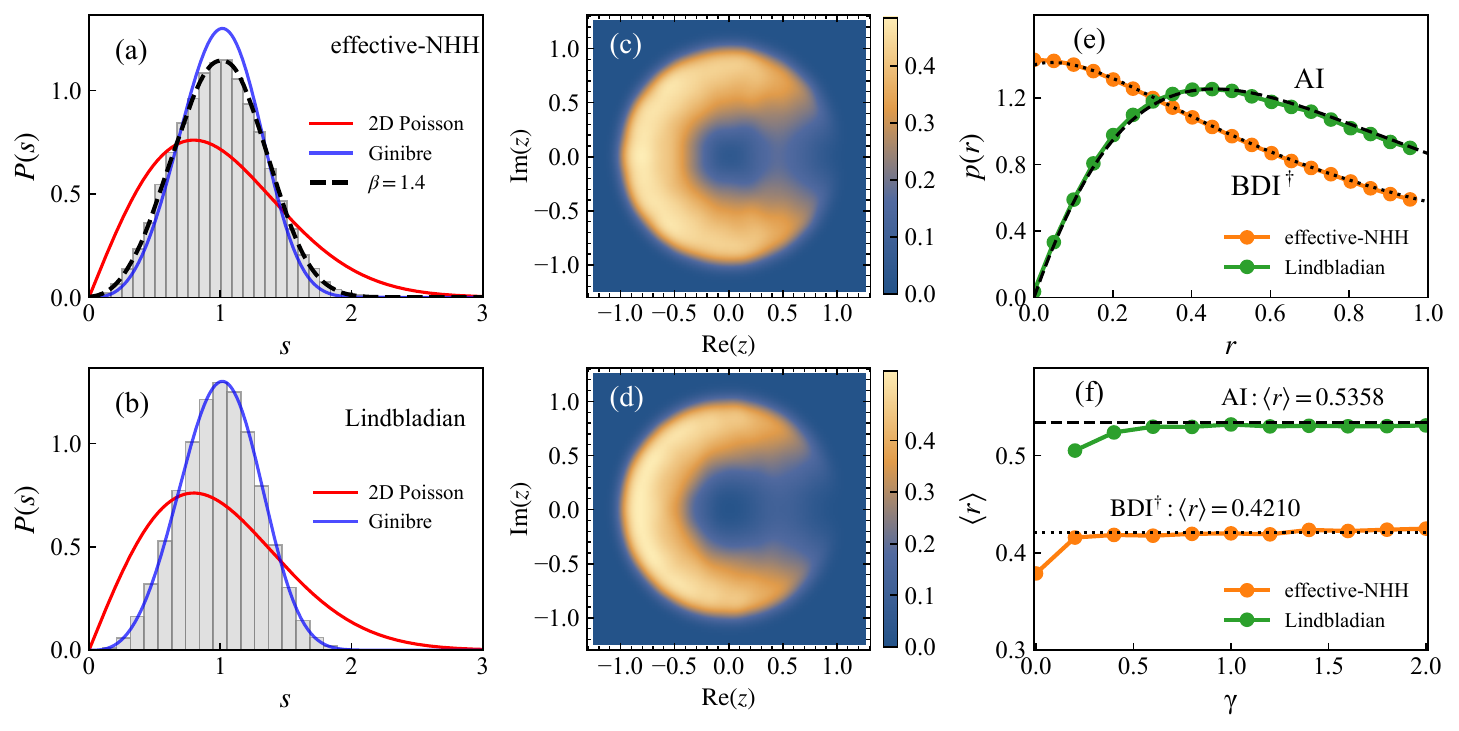}
    \caption{
        Spectral diagnostics for 
        the effective non-Hermitian Hamiltonian $H_{\rm eff}$ 
        and the Lindbladian $\mathbb L$ 
        in the dissipative transverse-field Ising chain with local damping.
        (a), (b) Nearest-neighbor level-spacing distribution $P(s)$ of the complex spectrum 
        for (a) $H_{\rm eff}$ and (b) $\mathbb L$, 
        compared with the two-dimensional (2D) Poisson distribution (red) and the Ginibre distribution (blue); 
        the dashed curve in (a) shows a two-dimensional Coulomb-gas reference at inverse temperature $\beta=1.4$.
        (c), (d) Distribution of complex spacing ratios (CSR) $z_n$ for (c) $H_{\rm eff}$ and (d) $\mathbb L$.
        (e) Singular-value spacing-ratio distribution $p(r)$; 
        the black dashed curves are the analytical results 
        for small non-Hermitian random matrices in classes AI and BDI$^{\dagger}$.
        (f) Mean ratio $\langle r\rangle$ versus dissipation strength $\gamma$, 
        with dashed (AI) and dotted (BDI$^{\dagger}$) lines. 
        For the singular-value statistics in (e), (f), 
        we discard $10\%$ of singular values at both spectral edges 
        to suppress finite-size edge effects. 
        (a)–(e) are computed at $\hbar\gamma=J$.
        The complete symmetry-resolved spectra contain $131\,328$ eigenvalues for $H_{\rm eff}$ and $130\,560$ eigenvalues for $\mathbb L$.
    }
    \label{fig:chaos_vs_chaos}
\end{figure*}

\section{Spectral correspondence and its breakdown}\label{sec:SecIII}

In this section, we perform a systematic comparison of the spectral statistics of 
Lindbladians $\mathbb{L}$ and their associated effective NHH $H_{\rm eff}$, 
focusing on when correspondence between the two generators is preserved or broken.
We begin with a regime where both generators exhibit Ginibre-like level repulsion, and
then turn to the 
Poisson regimes
to explore whether 
Poissonian statistics of
$H_{\rm eff}$ are inherited by the Lindbladian $\mathbb L$.
The complementary scenario, in which the Lindbladian exhibits Poissonian spectral statistics due to spectral separability, is discussed in the next section.

\subsection{Correspondence in the level-repulsive regime}\label{sec:SecIIIA}

We first illustrate a regime 
where the spectral diagnostics of the Lindbladian $ \mathbb{L} $ 
and the non-Hermitian Hamiltonian $ H_{\rm eff} $ 
are consistent and both exhibit Ginibre-like level repulsion.
As a representative realization, 
we consider a dissipative transverse-field Ising (TFI) chain 
with open boundary conditions,
\begin{align}
    H_\mathrm{TFI} =
    J \sum_{i = 1}^{L - 1} \sigma^{z}_{i} \sigma^{z}_{i+1} 
    + h \sum_{i= 1}^{L} \sigma^{x}_{i}, 
\label{eq:TFI}
\end{align}
where $ L $ is the number of spins, and $ \sigma_i^{x,z} $ denote Pauli operators acting on site $ i$.
The parameters $J$ and $h$ set the nearest-neighbor coupling strength 
and the transverse field, respectively.
This TFI model is integrable 
and can be mapped to free fermions via Jordan–Wigner transformation. 
In the thermodynamic limit, it exhibits a quantum phase transition 
at $ h = J $.
Dissipation is introduced via local damping jump operators,
\begin{align}
    L_{i} = \sqrt{\gamma_i} \sigma_{i}^{ -},
    \label{eq:disorder_damping}
\end{align}
where $\gamma_i$ denotes the local dissipation rate.
In this section, we consider uniform dissipation and set $\gamma_i=\gamma$
for all sites, with $\gamma$ controlling the overall dissipation strength.
For this choice of Hamiltonian and jump operators, 
both $\mathbb L$ (in the ladder representation) 
and $H_{\rm eff}$ respect the spatial reflection symmetry $P_x$.
In the following, we restrict the spectral analysis 
to the even-$P_x$ sector for both generators. 
We compute the complex spectra by exact diagonalization 
for $\mathbb L$ at $L=9$ and for $H_{\rm eff}$ at $L=18$.
In the largest system size considered, the Hilbert-space dimension is $131{,}328$, 
and we set $h=1.05J$ and $\hbar\gamma=J$.

We first analyze the nearest-neighbor level-spacing
distribution $P(s)$ of the complex eigenvalues for
$H_{\rm eff}$ and $\mathbb L$, shown in
Figs.~\ref{fig:chaos_vs_chaos}(a) and
\ref{fig:chaos_vs_chaos}(b), respectively.
For both generators, $P(s)$ exhibits clear level repulsion
and lies close to the GinUE distribution given in
Eq.~(\ref{eq:GinUE}) (blue curve).
From the correspondence perspective, the damping channel
contributes non-Hermitian terms to $H_{\rm eff}$ and
recycling processes to $\mathbb L$, which break the
free-fermion integrable structure and 
drive both spectra into a regime with Ginibre-like level repulsion.

For $H_{\rm eff}$, the complex-symmetry relation
$H_{\rm eff}^{T}=H_{\rm eff}$ places its bulk
complex-eigenvalue statistics in class
AI$^{\dagger}$, whose spacing distribution is effectively
described by a two-dimensional Coulomb gas in the range
$\beta\simeq1.3$--$1.4$~\cite{PhysRevE.106.014146,
PhysRevResearch.7.013098}.
As shown in Fig.~\ref{fig:chaos_vs_chaos}(a), the
$\beta=1.4$ reference (black dashed curve) provides an
excellent match to the numerical distribution, where the
Coulomb-gas reference curve is obtained by Monte Carlo
sampling of the joint distribution~\cite{2019JSP174692C,
2019PhRvL123y4101A}.
For $ \mathbb L $, the damping jumps do not
satisfy the transposition constraint. 
Therefore, its correlated complex-spectrum statistics 
are compared with the class A reference, represented by GinUE.

Additionally, we examine the distribution of the complex spacing ratios (CSR) 
$ z_n $ for $ H_{\rm eff} $ and $ \mathbb L $, 
shown in Figs.~\ref{fig:chaos_vs_chaos}(c) and \ref{fig:chaos_vs_chaos}(d), respectively. 
For both generators, the CSR distribution shows a pronounced suppression near the origin 
and a clearly anisotropic angular profile, 
consistent with level repulsion in the complex plane~\cite{2020PhRvX10b1019S}.
These features are characteristic of strongly correlated complex spectra
in contrast to the approximately uniform CSR expected for weakly correlated (Poisson) spectra.
The CSR therefore provides a 
consistent complex-spectrum correlation diagnostic
for both $ H_{\rm eff} $ and $ \mathbb L $, 
in agreement with the level-spacing distribution.

While $P(s)$ and the CSR consistently indicate strong
complex-eigenvalue correlations for both generators,
we next turn to singular-value statistics to probe 
whether this correspondence persists
at the level of symmetry classes.
Specifically, we compute the singular-value spacing ratios $r_n$ 
and the mean $\langle r\rangle$ for $H_{\rm eff}$ and $\mathbb L$, 
as depicted in Figs.~\ref{fig:chaos_vs_chaos}(e) and \ref{fig:chaos_vs_chaos}(f).
Fig.~\ref{fig:chaos_vs_chaos}(e) shows that 
the singular-value statistics of $\mathbb L$ belong to class AI, 
whereas those of $H_{\rm eff}$ are consistent with class BDI$^{\dagger}$~\cite{2023PRXQ4d0312K}.
Here, AI and BDI$^\dagger$ denote symmetry classes for 
the singular-value statistics of non-Hermitian matrices.
Class AI is characterized by an antiunitary
time-reversal symmetry $\mathcal T A\mathcal T^{-1}=A$
with $\mathcal T^2=+1$. Class BDI$^\dagger$ contains
TRS$^\dagger$ and PHS$^\dagger$ symmetries, both
squaring to $+1$, whose combination gives a chiral
symmetry $S A^\dagger S^{-1}=-A$.
These singular-value symmetry classes should be
distinguished from the bulk complex-eigenvalue class
AI$^\dagger$ discussed above.
This assignment is also corroborated by the mean ratio $\langle r\rangle$ 
in Fig.~\ref{fig:chaos_vs_chaos}(f), 
shown as a function of the dissipation strength $\gamma$.
We find that $\langle r\rangle$ rapidly approaches the class values 
$\langle r\rangle_{\rm AI}=0.5358$ for $\mathbb L$ 
and $\langle r\rangle_{\rm BDI^{\dagger}}=0.4210$ for $H_{\rm eff}$
~\footnote{As in many finite-size spectral diagnostics, convergence to the random-matrix class values typically improves with increasing matrix dimension (Hilbert-space for $H_{\rm eff}$ and Liouville-space for $\mathbb L$); correspondingly, the crossover in $\langle r\rangle$ may occur at smaller $\gamma$ for larger system sizes.}, 
and remains stable across the explored range of $\gamma$.
This is expected since $\mathbb L$ admits an antiunitary time-reversal symmetry 
(see Sec.~\ref{sec:SecII.B}), placing its singular-value statistics in class AI.
For $H_{\rm eff}$ considered here, 
the no-jump generator respects time-reversal symmetry$^\dagger $
\begin{align}
    H_{\rm eff}^{T} = H_{\rm eff}.
\end{align}
Moreover, $\text{i} H_{\rm eff}$ has a particle-hole symmetry$^\dagger $ 
\begin{align}
    \left( \prod_{i= 1}^{L} \sigma_{i}^{x} \right) (\text{i}  H_{\rm{eff}})^* \left( \prod_{i = 1}^{L} \sigma_{i}^{x} \right)^{- 1} = - \text{i}  H_{\rm{eff}}.
\end{align} 
The corresponding chiral symmetry follows from the combination of these two symmetries, 
placing the singular-value statistics of $H_{\rm eff}$ in class BDI$^{\dagger}$~\cite{2023PRXQ4d0312K}.
This demonstrates that 
even when both generators display Ginibre-like level repulsion,
their singular-value statistics can fall into different symmetry classes, 
revealing a finer-grained mismatch between $\mathbb L$ and $H_{\rm eff}$.

\subsection{Inheritance of Poisson spectral statistics}\label{sec:SecIIIC}

We now study
whether Poisson complex-spectrum statistics of 
the effective non-Hermitian Hamiltonian $H_{\rm eff}$ can be inherited by the full Lindbladian 
$\mathbb L$. 
Specifically, we consider an interacting \textit{XXZ} spin-$1/2$ chain and a non-interacting TFI chain,
where integrability stems from Bethe-ansatz and free-fermion structures respectively,
both coupled to an identical local dephasing channel.
This comparison allows us to examine how different integrable structures in $H_{\rm eff}$
influence the spectral statistics of the Lindbladian. 
For the dephasing Lindbladians considered here,
the relevant transposition symmetry is present, so that the
corresponding correlated random-matrix reference is class
AI$^\dagger$.

We first study an \textit{XXZ} spin-$ 1/2 $ chain of length $ L $ 
with open boundary conditions,
\begin{equation}
    H_\mathrm{XXZ} = 
        J \sum_{i = 1}^{L - 1} 
            \left( 
                \sigma^{x}_{i} \sigma^{x}_{i+1} 
                + \sigma^{y}_{i} \sigma^{y}_{i+1} 
                + \Delta \sigma^{z}_{i} \sigma^{z}_{i+1} 
            \right),
\label{eq:XXZ}
\end{equation}
where $J$ sets the nearest-neighbor exchange energy scale, 
and $\Delta$ controls the Ising anisotropy along the $z$ axis.
The \textit{XXZ} chain is Bethe-ansatz solvable and thus integrable.
Dissipation is introduced via local dephasing jump operators
\begin{align}
    L_i =\sqrt{\gamma}\sigma_i^z,
\end{align}
with $\gamma$ controlling the uniform dephasing strength in Eq.~(\ref{eq:masterEquation}).

For this choice of Hamiltonian and dephasing jumps, 
both $H_{\rm eff}$ and $\mathbb L$ 
possess a $U(1)$ symmetry associated with total $z$-magnetization, 
spatial reflection $P_x$, and a global $\mathbb Z_2$ spin-inversion symmetry.
In addition, $\mathbb L$ conserves the magnetization imbalance between the ket and bra legs, 
generated by the magnetization difference operator $ \mathcal{M}_d $ (see Sec.~\ref{sec:SecII.B}). 
Working in joint $ (m_p,m_d) $ sectors, 
the $\mathbb Z_2$ spin inversion maps $(m_p,m_d)\mapsto(-m_p,-m_d)$, 
and thus acts within a sector only when the selected sector is self-mapped.
In the following, we set $\Delta=1.05J$, and $\hbar\gamma=J$, 
and perform exact diagonalization for $\mathbb{L}$ on a chain of length $L=11$.
We focus on the even-$ P_x $ sector of $\mathbb L$ within the fixed $(m_p,m_d)=(2,0)$ block, 
where the global spin-inversion symmetry does not act within the sector.

\begin{figure}[tbp]
    \centering
    \includegraphics[width=1.0\linewidth]{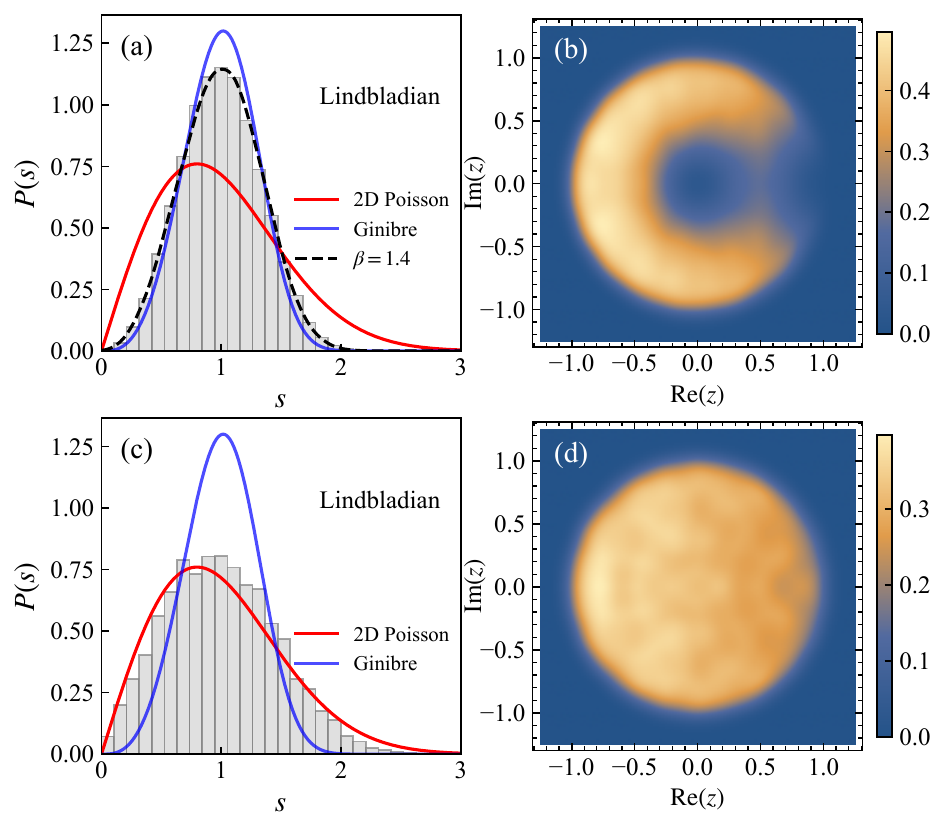}
    \caption{
        Spectral diagnostics for Lindbladians $\mathbb L$ in dephasing spin chains at $ \hbar\gamma = J $. Upper row: dephasing \textit{XXZ} chain; (a) nearest-neighbor level-spacing distribution $P(s)$, compared with the 2D Poisson statistics (red) and Ginibre distribution (blue), with a Coulomb-gas reference (black dashed) at $\beta=1.4$; (b) corresponding complex spacing ratio (CSR) distribution. Lower row: dephasing transverse-field Ising chain; (c) level spacing distribution $P(s)$; (d) corresponding CSR distribution. 
        The symmetry-resolved Lindbladian spectra contain $106{,}672$ eigenvalues for the dephasing \textit{XXZ} chain and $65{,}280$ eigenvalues for the dephasing TFI chain.
    }
    \label{fig:integer_vs_integer}
\end{figure}

In this dephasing setting,
the effective NHH $H_{\rm eff}$ reduces to the isolated
\textit{XXZ} Hamiltonian
up to an overall constant imaginary shift,
and therefore retains 
Poisson statistics.
For the Lindbladian, however, the dephasing channel generates Ising-type rung interactions
in the ladder picture, i.e., $ZZ$-type couplings between the ket and bra legs.
In the interacting \textit{XXZ} case, these rung interactions act as nontrivial couplings
and lead to 
strongly correlated complex spectra.
Specifically, Fig.~\ref{fig:integer_vs_integer}(a) shows that 
$P(s)$ for $\mathbb L$ 
is well described by the Coulomb-gas reference at $\beta=1.4$, consistent with class AI$^\dagger$.
This is further supported by Fig.~\ref{fig:integer_vs_integer}(b), 
where the CSR distribution displays a strongly nonuniform bulk pattern within the unit disk. 
These results provide a structural route to the 
coexistence of Poisson statistics for $H_{\rm eff}$ and level repulsion for $\mathbb L$:
for dephasing jumps, the non-Hermitian part of $H_{\rm eff}$ is only an overall imaginary shift, 
while $\mathbb L$ exhibits complex-spectrum level repulsion.

We now turn to the dephasing channel acting on a free-fermion integrable Hamiltonian,
namely the TFI chain introduced in Eq.~\eqref{eq:TFI}.
For this choice of Hamiltonian and jump operators, 
both $\mathbb L$ and $H_{\rm eff}$ respect the spatial reflection symmetry $P_x$ 
and the spin-inversion $\mathbb Z_2$ symmetry.
In the following calculations, we set $h=1.05J$ and $\hbar\gamma=J$ 
for $\mathbb{L}$ on a chain of length $L=9$
and focus on the sector with even-$\mathbb Z_2$ parity and even-$P_x$ parity.

In this dephasing setting, $H_{\rm eff}$ also remains 
free-fermionic
since its spectrum differs from that of the isolated TFI Hamiltonian
only by an overall imaginary shift.
For the full Lindbladian, the lower panels of Fig.~\ref{fig:integer_vs_integer} show that the spectral diagnostics are consistent with 
weakly correlated complex-spectrum statistics.
Specifically, the complex spacing distribution $P(s)$ follows the 2D Poisson form [Fig.~\ref{fig:integer_vs_integer}(c)], 
and the CSR distribution is close to uniform within the unit disk [Fig.~\ref{fig:integer_vs_integer}(d)].
In contrast to the interacting \textit{XXZ} chain,
the dephasing-induced rung interactions in the Lindbladian
do not induce 
level repulsion
in this open non-interacting TFI model.

These two dephasing settings constitute a controlled structural comparison under an identical dissipative channel.
While the interacting \textit{XXZ} chain has an effective NHH $H_{\rm eff}$ that remains integrable, the Lindbladian $\mathbb L$ exhibits 
level repulsion.
By contrast, for the free-fermion transverse-field Ising chain, both $H_{\rm eff}$ and $\mathbb L$ display weakly correlated (Poisson) complex-spectrum statistics.
These results demonstrate that, 
for the dephasing channel, 
whether Poisson spectral statistics of $H_{\rm eff}$ are inherited by the Lindbladian depends sensitively on the underlying coherent Hamiltonian, and is not determined by the dissipative channel alone.

\section{Poissonian Lindbladian spectra from structural constraints}\label{sec:SecIV}

In this section, we address the complementary scenario announced above: 
an effective NHH $H_{\rm eff}$ with 
strongly correlated statistics
coexisting with Poissonian complex-spectrum statistics of the Lindbladian $\mathbb L$.
We show that this behavior can arise from a broad class of Lindbladians 
whose spectra admit a \textit{separable structure} in Liouville space, 
such that the eigenvalues of $\mathbb L$ are exactly constructible 
from those of $H_{\rm eff}$,
while their spectral correlations are substantially modified.

We begin by introducing this class of spectrally separable Lindbladians 
at a general structural level.
Then, we consider a concrete realization with disordered dissipation, 
which completes our classification 
by demonstrating that Poissonian Lindbladian statistics persist
even when $H_{\rm eff}$ displays 
strongly correlated complex-spectrum statistics.
Finally, we turn to uniform dissipation as a particularly clean and transparent realization,
in which the complex Lindbladian spectrum displays Poissonian real-spacing statistics
and exhibits a characteristic banded structure in Liouville space.

\subsection{Spectrally separable Lindbladians}\label{sec:U1_damping_triangular}

In the vectorized (Liouville-space) representation, a general Lindbladian can be written as
\begin{equation}
    \mathbb{L} = \mathbb{H}_{\text{eff}} + \mathbb{R},
\end{equation}
where we introduce the Liouville-space superoperators
\begin{equation}
    \mathbb{H}_{\rm eff} \equiv -\mathrm{i}\big(H_{\rm eff}\otimes \mathbb{I}
    - \mathbb{I} \otimes H_{\rm eff}^*\big),
    \;\;
    \mathbb R \equiv \sum_{i} L_{i} \otimes L^*_{i},
\end{equation} 
representing the no-jump and recycling contributions.

We consider systems in which the effective non-Hermitian Hamiltonian $H_{\rm eff}$
conserves a charge $Q$ defined on the physical Hilbert space, i.e.,
\(
    [H_{\rm eff},Q]=0.
\)
This condition implies that the associated no-jump superoperator
$\mathbb{H}_{\text{eff}}$ preserves the total charge, so that
\([\mathbb{H}_{\text{eff}},\mathcal Q]=0,
\)
where $\mathcal Q = Q_\text{ket} + Q_\text{bra}$ labels the sum of the charges in the ket and bra spaces.
As a result, $\mathbb{H}_{\text{eff}}$ is block diagonal in a basis ordered by $\mathcal Q$.

In contrast, we further consider systems in which the jump operators $L_i$ 
act in a strictly one-sided manner on the charge sectors of $\mathcal Q$,
i.e., either lowering or raising the charge.
In what follows, we focus on the lowering case.
If $L_i$ lowers the charge by one unit, the recycling term $\mathbb R$ lowers the total Liouville-space charge by two units.
Consequently, $\mathbb R$ connects only sectors labeled by the eigenvalue $q$ of $\mathcal Q$ with $ q\to q-2$, and therefore appears exclusively in off-diagonal blocks.

\begin{figure}[t]
  \centering
  \begin{tikzpicture}[x=1.5cm,y=1.25cm] 
    \def\n{4}        
    \def\s{1.1}      
    \def\pad{0.35}   

    \coordinate (LL) at (-\pad, -\pad);
    \coordinate (UR) at (\n*\s+\pad, \n*\s+\pad);

    \draw[line width=0.7pt] (LL) -- (-\pad, \n*\s+\pad);
    \draw[line width=0.7pt] (-\pad, \n*\s+\pad) -- (0, \n*\s+\pad);
    \draw[line width=0.7pt] (LL) -- (0, -\pad);

    \draw[line width=0.7pt] (\n*\s+\pad, -\pad) -- (\n*\s+\pad, \n*\s+\pad);
    \draw[line width=0.7pt] (\n*\s+\pad, \n*\s+\pad) -- (\n*\s, \n*\s+\pad);
    \draw[line width=0.7pt] (\n*\s+\pad, -\pad) -- (\n*\s, -\pad);

    \foreach \k/\lab in {0/{q-2},1/{q-1},2/{q},3/{q+1}}{
      \pgfmathsetmacro{\x}{\k*\s}
      \pgfmathsetmacro{\y}{(\n-1-\k)*\s}
      \draw[line width=0.7pt] (\x,\y) rectangle ++(\s,\s);
      \node[font=\small] at (\x+0.5*\s,\y+0.5*\s) {$\mathbb{H}_{\text{eff}}^{(\lab)}$};
    }

    \foreach \i in {0,1}{
    \pgfmathsetmacro{\x}{(\i+2)*\s}
    \pgfmathsetmacro{\y}{(\n-1-\i)*\s}
    \draw[line width=0.7pt] (\x,\y) rectangle ++(\s,\s);

    \ifnum\i=0
        \node[align=center, font=\small] at (\x+0.5*\s,\y+0.5*\s)
        {$\mathbb R_{q}^{q-2}$};
    \else
        \node[align=center, font=\small] at (\x+0.5*\s,\y+0.5*\s)
        {$\mathbb R_{q+1}^{q-1}$};
    \fi
    }

    \node[font=\large] at (-0.1*\s, \n*\s+0.17) {$\cdots$};
    \node[font=\large] at (-0.1*\s+2.2, \n*\s+0.17) {$\cdots$};
    \node[font=\large] at (\n*\s+0.1*\s, -0.15*\s) {$\cdots$};
    \node[font=\large] at (\n*\s+0.1*\s, -0.15*\s+2.2) {$\cdots$};

  \end{tikzpicture}

  \caption{
    Schematic block-triangular structure of $\mathbb{L} = \mathbb{H}_{\text{eff}} + \mathbb{R}$ in a basis ordered by the total Liouville-space charge $\mathcal Q$.
    The no-jump superoperator $\mathbb{H}_{\text{eff}}$ is block diagonal with blocks $\mathbb{H}_{\text{eff}}^{(q)}$ acting within fixed-$q$ sectors.
    For jump operators that lower the physical charge by one unit, the recycling term $\mathbb{R}=\sum_i L_i\otimes L_i^*$ connects only sectors $q \to q-2$, yielding nonzero off-diagonal blocks $\mathbb{R}^{q-2}_{q}$.
    Block sizes are schematic and not drawn to scale.
  }
  \label{fig:block_triangular_Mp}
\end{figure}
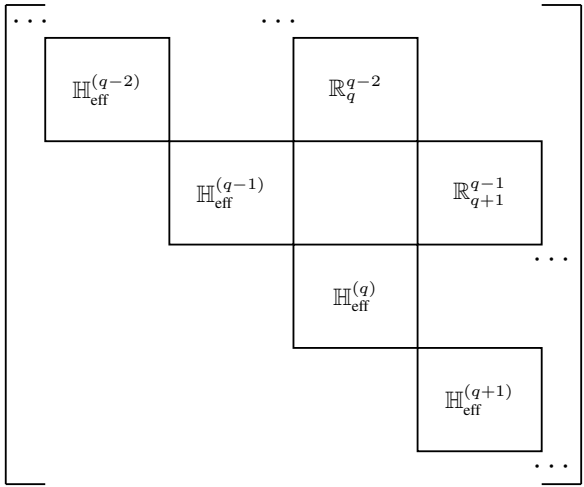

In the $\mathcal Q$-ordered basis, the full Lindbladian acquires a block-triangular structure, with diagonal blocks given by those of $\mathbb H_{\rm eff}$ (see Fig.~\ref{fig:block_triangular_Mp}), which we refer to as $\mathbb H_{\rm eff}^{(q)}$.
A direct consequence is that the eigenvalue set of $\mathbb L$ is completely determined by its diagonal blocks, i.e.,
\begin{equation}
    \mathrm{spec}(\mathbb L)
    =
    \mathrm{spec}(\mathbb H_{\rm eff})
    =
    \bigcup_{q}\mathrm{spec}\!\left(\mathbb H_{\rm eff}^{(q)}\right),
\label{eq:spec_L_equals_Heff}
\end{equation}
More explicitly, if $E_\mu$ denotes an eigenvalue of
$H_{\rm eff}$, the spectrum of the no-jump superoperator $\mathbb H_{\rm eff}$
is given by pairwise differences of these eigenvalues,
\begin{equation}
    \lambda_{\mu\nu} = -\,\mathrm{i}\,(E_\mu - E_\nu^{*}).
\end{equation}
This decoupled difference structure explicitly shows that the Lindbladian spectrum
is constructed from two independent copies of the $H_{\rm eff}$ spectrum.
We refer to this construction as \emph{spectrally separable}. 
Related triangular structures 
for loss-only Lindbladians with a conserved charge 
have also been discussed~\cite{2014PhysRevA89052133,Buca2020NJPLossBethe,2021PhysRevLett126110404}, 
primarily in the context of exact spectral constructions and relaxation properties.


It is important to emphasize that this spectral separability
does not mean a symmetry decomposition of the Lindbladian.
It is also distinct from the ordinary charge-sector
decomposition of $H_{\rm eff}$ in the physical Hilbert
space, where spectral statistics can be evaluated within a
fixed charge sector. In the Liouville-space construction
above, the diagonal $q$ blocks of $\mathbb H_{\rm eff}$ are
not independent symmetry sectors of the full Lindbladian,
because the recycling term $\mathbb R$ connects different
$q$ sectors. Although the eigenvalues of $\mathbb L$ are
therefore completely determined by the diagonal blocks of
$\mathbb H_{\rm eff}$, the off-diagonal recycling terms mix
different charge sectors at the level of eigenmodes. As a
result, the relevant Lindbladian spectrum is the union of
the diagonal-block spectra rather than the spectrum of a
single block. This pairwise-difference and sector-superposition
structure provides a mechanism for suppressing level
repulsion, as demonstrated numerically below.

We now turn to concrete realizations of the spectrally separable structure discussed above.
A particularly natural setting is provided by spin chains with conserved total magnetization.
The coherent Hamiltonian $H$ preserves the charge of 
$Q=\sum_i \sigma_i^z$,
while dissipation is implemented through local spin-lowering jump operators, as defined in Eq.~\eqref{eq:disorder_damping}, with site-dependent rates $\gamma_i$.
These choices ensure that the effective NHH
$H_{\rm eff}$ conserves the charge $Q$, whereas the recycling term changes the
corresponding Liouville-space charge in a strictly one-sided manner.

\subsection{Disordered dissipation realization}

We first consider the case of disordered dissipation, $L_i=\sqrt{\gamma_i}\,\sigma_i^{-}$, where the damping rates
$\gamma_i$ are taken to be site-dependent.
As a concrete realization, we choose the coherent Hamiltonian $H$ to be the
interacting \textit{XXZ} chain introduced in Eq.~(\ref{eq:XXZ}).
Disorder is implemented by sampling the rates $\{\gamma_i\}$ independently
from a uniform distribution $\gamma_i\in[0,\gamma]$, where $\gamma$ controls
the disorder strength.
For $H_{\rm eff}$, we focus on the zero-magnetization sector ($m_p=0$) at chain length $L=18$.
For the Lindbladian $\mathbb L$, we consider a chain of length $L=9$
and restrict to the $m_d=0$ symmetry sector.
Throughout, we set $\Delta=J$.
The disordered effective NHH remains complex symmetric
and therefore has the class AI$^\dagger$ correlated
random-matrix reference, 
whereas the damping Lindbladian
itself lacks the TRS$^\dagger$ transposition symmetry.

\begin{figure}[tbp]
    \centering
    \includegraphics[width=1.0\linewidth]{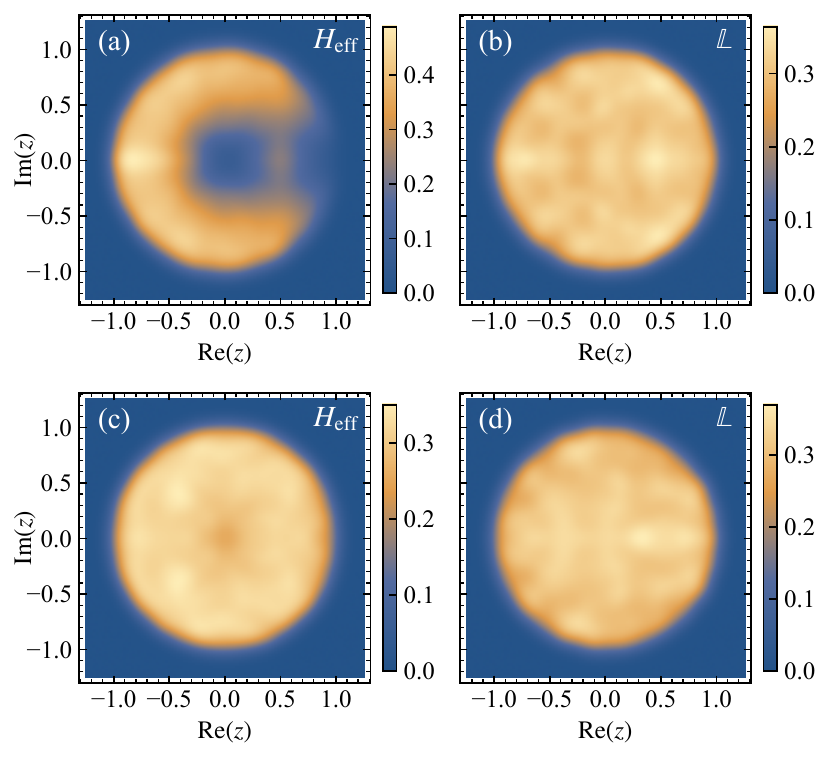}
    \caption{
        Complex-spacing-ratio (CSR) distributions 
        for the effective NHH $H_{\rm eff}$ 
        and the Lindbladian $\mathbb{L}$ 
        in the disordered damped \textit{XXZ} chain. 
        (a) and (c) show $H_{\rm eff}$ at different disorder strengths, 
        $\hbar\gamma=2J$ and $20J$, respectively, 
        while panels (b) and (d) show the corresponding results for $\mathbb L$. 
        For each disorder strength, the statistics are averaged over $10$ independent disorder realizations, with $48{,}620$ eigenvalues in each symmetry-resolved spectrum for both $H_{\rm eff}$ and $\mathbb L$.
    }
    \label{fig:MBL}
\end{figure}

Fig.~\ref{fig:MBL} shows the CSR distributions for the effective
NHH $H_{\rm eff}$ [Figs.~\ref{fig:MBL}(a), (c)] and for the
Lindbladian $\mathbb L$ [Figs.~\ref{fig:MBL}(b), (d)] at $\gamma=2J$ and
$\gamma=20J$, respectively.
For $\gamma=2J$, the CSR distribution of $H_{\rm eff}$ is strongly nonuniform
[Fig.~\ref{fig:MBL}(a)], consistent with 
strongly correlated complex-spectrum statistics.
Here, level repulsion arises from the presence of disordered dissipation, which 
breaks the Bethe-ansatz structure that is present for uniform damping.
By contrast, the Lindbladian spectrum exhibits a nearly uniform CSR
distribution [Fig.~\ref{fig:MBL}(b)], indicating 
approximately uncorrelated complex spectra
despite
the strongly correlated CSR
of $H_{\rm eff}$.
This behavior reflects the fact that 
the spectral statistics of the Lindbladian
are governed by the underlying spectrally separable Liouville-space structure,
rather than by the detailed spectral correlations of $H_{\rm eff}$. 
More concretely, within the fixed Lindbladian symmetry
sector considered here, the eigenvalues entering the
statistics form a superposition of the diagonal-block spectra
appearing in the triangular construction discussed above.
This superposition of pairwise-difference spectra suppresses
the level repulsion visible in $H_{\rm eff}$, leading to the
nearly uniform CSR distribution of $\mathbb L$.

Upon increasing the disorder strength to $\gamma=20J$, 
the CSR distribution of $H_{\rm eff}$ 
becomes much closer to a uniform disk 
within the system sizes accessible here [Fig.~\ref{fig:MBL}(c)], 
indicating, for this system size, strongly reduced spectral correlations and a Poisson-like regime. 
A similar change has been reported for the same non-Hermitian Hamiltonian 
in Ref.~\cite{PhysRevB109L140201}, where singular-value based diagnostics provide evidence for a localization-like regime at strong dissipation disorder. We emphasize that a systematic finite-size scaling would clarify the nature of this crossover/transition, but this is beyond the scope of this work. 
In contrast, the Lindbladian's CSR distribution remains essentially unchanged
[Fig.~\ref{fig:MBL}(d)], further illustrating the robustness of Poissonian statistics for the Lindbladian within the $U(1)$-symmetric damping family. 

\subsection{Uniform damping and band spectral structure}

To gain further insight into the consequences of spectral separability in
Liouville space, we now turn to a particularly clean realization within the
$U(1)$-symmetric damping family.
In this setting, dissipation is taken to be uniform, $L_i=\sqrt{\gamma}\,\sigma_i^{-}$,
while the coherent Hamiltonian $H$ itself is nonintegrable
and exhibits chaotic spectral statistics. 
We illustrate this structure using a next-to-nearest-neighbor (NNN) Heisenberg \textit{XXZ} chain with Hamiltonian
\begin{align}
    H &=
        J \sum_{i=1}^{L-1}
            \!\left(
                \sigma_i^x \sigma_{i+1}^x
                + \sigma_i^y \sigma_{i+1}^y
                + \Delta\,\sigma_i^z \sigma_{i+1}^z
            \right) \nonumber \\
    &\quad +
        \ J' \sum_{i=1}^{L-2}
            \!\left(
                \sigma_i^x\sigma_{i+2}^x
                + \sigma_i^y\sigma_{i+2}^y
                + \Delta'\,\sigma_i^z\sigma_{i+2}^z
            \right),
\label{eq:NNN_XXZ}
\end{align}
where $J$ ($J'$) denotes the nearest- (next-to-nearest-) neighbor exchange coupling, 
and $\Delta$ $(\Delta')$ denotes the corresponding $z$-axis anisotropy.
Dissipation is implemented via uniform damping rates $\gamma_i\equiv\gamma$ [Eq.~\eqref{eq:disorder_damping}].
We work in the zero-magnetization sector for $H_{\rm eff}$.
Within this sector, two additional symmetries, 
spin inversion and spatial reflection ($P_x$), are present, 
and we focus on the even inversion and even-$P_x$ sector.
For the Lindbladian $\mathbb L$, 
the magnetization difference $\mathcal{M}_d$ is conserved, 
and we restrict to $m_d=0$ and even-$P_x$ sector.
We set $J'=J, \Delta=0.5 J, \Delta'=1.5 J$, 
and consider chains of length $L=10$ for $\mathbb{L}$ and $L=20$ for $H_\text{eff}$ with damping strength $ \hbar\gamma=J $.

\begin{figure}[tbp]
    \centering
    \includegraphics[width=1.0\linewidth]{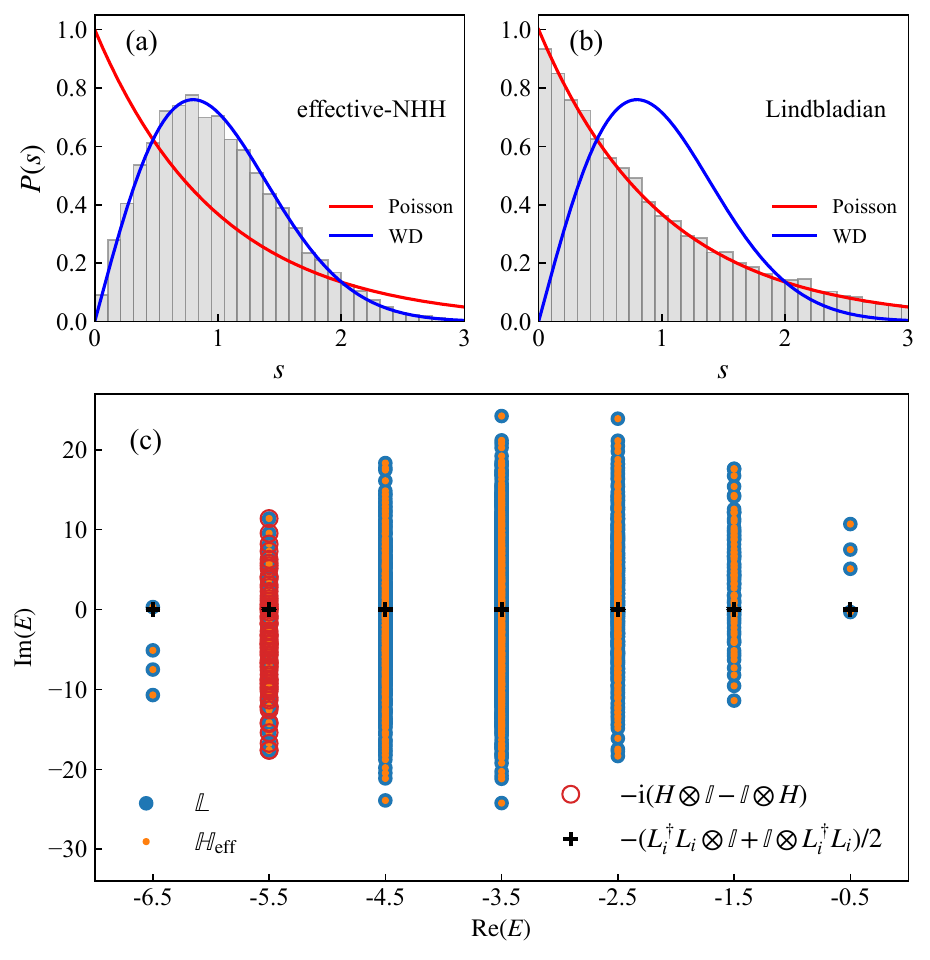}
    \caption{Spectral diagnostics for the next-to-nearest-neighbor Heisenberg \textit{XXZ} chain with uniform damping. (a), (b) Nearest-neighbor level-spacing distribution $P(s)$ for $H_{\rm eff}$ and for $\mathbb L$, respectively; the red and blue curves are the Poisson and Wigner-Dyson reference distributions for real-spectrum level-spacing statistics. (c) Complex spectrum of $\mathbb L$ (blue dots) and of the no-jump superoperator $\mathbb H_{\rm eff}$ (orange dots). Black crosses indicate the damping shifts from $-(L^\dagger_{i} L_{i}\otimes\mathbb I+\mathbb I\otimes L^\dagger_{i} L_{i})/2$, and the red circles show the spectrum of the coherent part $-\mathrm{i}(H\otimes\mathbb I-\mathbb I\otimes H)$ in the $(m_{\rm ket}=5,m_{\rm bra}=3)$ block (with $m_d=2$). 
    The symmetry-resolved spectra used in (a) and (b) contain $46{,}508$ eigenvalues for $H_{\rm eff}$ and $92{,}252$ eigenvalues for $\mathbb L$, respectively.
    }
    \label{fig:J1J2}
\end{figure}

Although both $H_{\rm eff}$ and the Lindbladian $\mathbb L$ are
non-Hermitian operators, their level-spacing statistics are well described by the familiar distributions of
Hermitian systems.
As shown in Fig.~\ref{fig:J1J2}(a), the level-spacing distribution
$P(s)$ of $H_{\rm eff}$ agrees well with the Wigner--Dyson form,
$P(s) = (\pi s/2)\exp(-\pi s^{2}/4)$~\cite{1998PhR299189G,1977RSPSA356375B},
indicating statistics consistent with the Gaussian orthogonal
ensemble (GOE).
This behavior follows directly from the structure of $H_{\rm eff}$.
For uniform jumps, the non-Hermitian contribution to $H_{\rm eff}$
is proportional to $\sum_i L_i^\dagger L_i$, which reduces to
$\gamma(M+L)/2$, where
$M=\sum_i\sigma_i^z$ is the conserved magnetization and $L$ is the
chain length.
It therefore contributes only a constant within each
fixed-magnetization sector.
Accordingly, the spectral statistics of $H_{\rm eff}$ coincide with
those of the underlying nonintegrable NNN Heisenberg Hamiltonian,
displaying GOE statistics in the relevant symmetry sector.

In contrast,  
the level spacing distribution $P(s)$ of the Lindbladian
shown in Fig.~\ref{fig:J1J2}(b) is well described by the Poisson distribution,
$P(s)=e^{-s}$~\cite{1991PhRvA434237P,2013PhRvE88c2913Z,2018NJPh20k3039G},
exhibiting the uncorrelated spacing behavior familiar from
integrable Hermitian systems, despite the fact that $\mathbb L$ 
possesses a generically complex spectrum.

The essential ingredient for understanding this behavior is that the coherent Hamiltonian $H$
commutes with the non-Hermitian contribution,
\begin{equation}
    [H,\; \sum_i L_i^\dagger L_i] = 0,
\end{equation}
which implies that the two operators can be
simultaneously diagonalized.
Since the no-jump superoperator can be written as
\(
\mathbb H_{\rm eff}
=
- \mathrm{i}(H\otimes \mathbb{I} - \mathbb{I}\otimes H)
- \frac{\gamma}{4}(\mathcal M_p + 2L),
\)
this simultaneous diagonalizability lifts directly to Liouville
space.
As a consequence, the eigenvalues of $\mathbb H_{\rm eff}$ take the simple form
\begin{equation}
    \lambda_{\alpha\beta}^{(m_p)}
    =
    -\mathrm{i}(E_\alpha - E_\beta)
    - \frac{\gamma}{4}(m_p + 2L),
\end{equation}
where $E_\alpha$ and $E_\beta$ are eigenenergies of $H$, and $m_p$
denotes the total magnetization of the corresponding Liouville
state.
As discussed in Sec.~\ref{sec:U1_damping_triangular}, 
these eigenvalues coincide with those of the Lindbladian $\mathbb L$.
The coherent part therefore generates purely imaginary eigenvalues
given by energy differences, while the damping term contributes only
a constant real shift controlled by $m_p$.

Since the Lindbladian $\mathbb L$ conserves only the magnetization
difference $\mathcal M_d$, its spectrum must be analyzed within fixed
$m_d$ sectors.
Within a given $m_d$ sector, the total magnetization $m_p$ is not
conserved and can take only a discrete set of allowed values.
As a result, the real parts of the eigenvalues,
$-\frac{\gamma}{4}(m_p+2L)$, form a discrete and equally spaced set,
giving rise to a characteristic banded structure in the complex
plane.
Within each band, the eigenvalues originate from energy differences
of the underlying coherent Hamiltonian and are expected to be only weakly
correlated.
Nearest neighbors in the complex plane are therefore predominantly
drawn from within the same band, leading to effectively uncorrelated
real-part spacings and hence Poisson statistics. 

Fig.~\ref{fig:J1J2}(c) provides a direct numerical confirmation of this picture.
The eigenvalues of the Lindbladian $\mathbb L$ (blue dots) coincide with those of $\mathbb H_{\rm eff}$ (orange dots), consistent with the block-triangular structure discussed in Sec.~\ref{sec:U1_damping_triangular}.
The red circles and black crosses correspond to the two commuting
contributions $-\mathrm{i}(H\otimes \mathbb{I} - \mathbb{I}\otimes H)$ and $-(\gamma/4)(\mathcal M_p+2L)$ appearing in $\mathbb H_{\rm eff}$, respectively, illustrating how the two terms combine additively in the spectrum. 
Because $m_p$ takes discrete values within a fixed-$m_d$ sector, 
the real shifts form an equally spaced set, 
and each vertical band is obtained by adding the imaginary energy-difference spectrum to one such shift.
This construction explains the band centers and their vertical extent, 
and it also explains why the nearest-neighbor spacings are dominated by intra-band pairs.

In the uniform-dissipation setting considered here, the analysis above
makes it explicit how the separable structure of Liouville space
organizes the spectrum.
The coherent Hamiltonian, the non-Hermitian damping terms, and the
recycling processes contribute in a clearly distinguishable manner, 
giving rise to a rigid banded structure and
Poissonian real-spacing statistics, despite the spectrum being
genuinely complex.
In this way, the uniform-dissipation case provides a particularly
transparent illustration of Poissonian statistics in Lindbladians arising
from spectral separability. For completeness, we note that a similar banded (or stripe-like) structure has also been reported in Ref.~\cite{Zundel2025} for driven-dissipative hard-core bosons with both local pump and loss. In that case, however, the bands arise only perturbatively in the weak-dissipation regime, providing a complementary perspective on how deviations from strictly one-sided dissipation (e.g. only losses or pumping) restore nonintegrable non-Hermitian spectral behavior.

\section{Summary}\label{sec:SecV}
In this work, we investigate how spectral statistics of Lindbladians $\mathbb{L}$ relate to those of their associated no-jump effective non-Hermitian Hamiltonians $H_{\rm eff}$.
Using a ladder representation and symmetry-resolved complex-spectrum diagnostics, 
we systematically characterize 
Poisson versus 
strongly correlated complex-spectral statistics
of both $\mathbb{L}$ and $H_{\rm eff}$.
Notably, we realize all combinations of 
weakly and strongly correlated spectral statistics between
effective non-Hermitian Hamiltonians and Lindbladians in paradigmatic dissipative spin chains, while also uncovering a broad class of spectrally separable Lindbladians whose complex spectra exhibit robust Poisson statistics.

We first show that dissipation drives both $\mathbb L$ and $H_{\rm eff}$ into a 
Ginibre-like regime with closely matching spectral statistics in the locally damped transverse-field Ising chain.
We then explore whether 
Poisson complex-spectrum statistics
of $H_{\rm eff}$ are necessarily inherited by the full Lindbladian $\mathbb L$.
Comparing an \textit{XXZ} chain with a transverse-field Ising chain, we find that the inheritance is highly sensitive to 
the coherent Hamiltonian.
Specifically, with identical dephasing jump operators, the interacting \textit{XXZ} model shows 
complex-spectrum level repulsion in the Lindbladian,
whereas the free-fermion transverse-field Ising chain retains 
Poisson statistics for both $\mathbb L$ and $H_{\rm eff}$.

Finally, we identify a broad structural family of Lindbladians whose Liouville-space spectra are separable and can be constructed exactly from the spectrum of the associated $H_{\rm eff}$.
Within this family, the Lindbladian generically exhibits Poissonian spectral statistics, even when $H_{\rm eff}$ displays 
strongly correlated complex-spectrum statistics
or undergoes a disorder-driven change toward Poisson-like behavior.
In the special case of uniform damping, we further uncover a characteristic phenomenology in which the level statistics resemble those of a real-spectrum Poisson ensemble, despite the Lindbladian possessing a genuinely complex spectrum.
This illustrates how structural constraints in Liouville space can suppress level repulsion, yielding Poissonian complex-spectrum statistics despite strongly correlated spectral signatures in the associated effective NHH.

These examples also clarify what can and cannot be predicted
from the no-jump Hamiltonian alone. 
Neither the integrability of $H_{\rm eff}$ nor its
random-matrix statistics alone provides a sufficient 
predictor of the Lindbladian statistics. 
A useful structural diagnostic is to first resolve the relevant symmetries and then examine how the recycling term acts in Liouville
space. 
Recycling-induced inter-leg couplings can generate
level repulsion in $\mathbb L$ even when $H_{\rm eff}$ is
Poissonian, 
whereas a spectrally separable block-triangular structure can organize the Lindbladian spectrum through pairwise-difference block superposition and thereby support Poissonian Lindbladian statistics even when $H_{\rm eff}$ is strongly correlated.
Thus, correspondence or breakdown between $\mathbb L$ and $H_{\rm eff}$ is governed by the compatibility between the no-jump dynamics, recycling,
and Liouville-space symmetry structure.

Building on this structural viewpoint, several open directions naturally follow from our work.
First, it remains to be clarified what dynamical information is encoded 
in complex-spectrum statistics of Lindbladians and their associated effective NHH, 
and how this depends on recycling processes and Liouville-space structure~\cite{Villasenor2024,Villasenor2025,Mondal2026}.
A second direction is to explore how the spectral separability
extends to more general dissipative channels, and to clarify its dynamical
implications, including relaxation~\cite{PhysRevE92042143}, thermalization~\cite{PhysRevB100045112,2025PhysRevLett134180405,qy19nc4r} and localization~\cite{PhysRevLett123090603, XuPoletti2018, VakulchykDenisov2018, XuPoletti2021} behavior.
More broadly, it would be interesting to connect these questions to wider classes of non-Hermitian and driven open systems, 
such as models with gain and loss competition~\cite{2007RPPh70947B,2019PhysRevX9041015}, Liouvillian exceptional points~\cite{PhysRevA100062131}, or Floquet Lindbladians~\cite{PhysRevE90012110}. 

\section{Acknowledgments} 
We warmly thank Lucas Sá for helpful advice and correspondence, and David Villaseñor and Lea F. Santos for helpful discussions. 
D.W. and D.P. acknowledge support from the Singapore Ministry of Education grant MOE-T2EP50123-0017, and from the Centre for Quantum Technologies grant CQT$\_$SUTD$\_$2025$\_$01. H.Z. and GF.Z. acknowledge support from the National Natural Science Foundation of China (Grant No. 12474353).

\bibliography{main.bib}

\end{document}